\documentclass[12pt]{iopart}
\usepackage{graphicx}  
\usepackage{xcolor}
\usepackage{soul}

\usepackage{iopams}
\expandafter\let\csname equation*\endcsname\relax
\expandafter\let\csname endequation*\endcsname\relax
\usepackage{amsmath}
\usepackage{cite}
\begin{document}

\title[ Density of states of a  $2 {\rm D} $ system]
{Density of states of a 2D system of soft–sphere fermions by path integral Monte Carlo simulations}
\author{V. Filinov$^1$,  P. Levashov$^{1,2}$, A. Larkin$^1$ } 

\address{$^1$Joint Institute for High Temperatures, Russian Academy of Sciences, \\
	Izhorskaya 13 bldg 2, Moscow 125412, Russia\\
	$^2$Moscow Institute of Physics and Technology, 9 Institutskiy per., \\ Dolgoprudny, Moscow Region, 141700, Russia}
\ead{vladimir\_filinov@mail.ru}
\vspace{10pt}
\begin{indented}
\item[]April 2023
\end{indented}

\begin{abstract}
The Wigner formulation of quantum mechanics is used to derive a new path integral representation of
quantum density of states. A path integral Monte Carlo approach is developed for the numerical investigation of
density of states, internal energy and spin--resolved radial distribution functions for a
2D system of strongly correlated soft--sphere fermions.
The peculiarities of the density of states and internal energy distributions depending on  
the hardness of the soft--sphere potential and particle density are investigated and explained. In particular, at high enough densities the density of states rapidly tends to a constant value, as for an ideal system of 2D fermions.
\end{abstract}

%
\submitto{\JPA}
%
%
%

\section{Introduction}
Density of states (DOS) is a key factor in condensed matter physics determining many properties of matter
\cite{harrison2012electronic}. The DOS is proportional to the fraction of states per unit volume that have a certain energy.
The product of the DOS and the probability distribution function gives the fraction  of occupied states
at a given energy for a system in thermal equilibrium.
Computing DOS is of fundamental importance
and many works for different systems are devoted to this problem.
Popular approaches are based on generalized ensembles and reweighting techniques. 
One of the most prominent approaches is the Wang--Landau (WL) algorithm, which
is a well-known Monte Carlo technique for computing the DOS of classical systems
\cite{wang2001determining,wang2001efficient,faller2003density,vogel2013generic,liang2005generalized,moreno2022portable,bornn2013adaptive,atchade2010wang}.  
 
DOS is much more important for determining the properties of quantum systems of particles \cite{martin2004electronic}.
For instance, researchers working in the solid--state and condensed matter physics usually apply quantum mechanical approaches, such as density functional theory (DFT) \cite{seo2014first,ma2015machine}.
However, DFT is approximate and in some cases leads to a severe computational workload \cite{ratcliff2017challenges}.

Thus, there are many attempts to develop fast and high--accuracy methods to predict the DOS and internal energy
distribution of materials \cite{galli1996linear,saad2010numerical,goedecker1999linear}.
An interesting approach involving path integrals was suggested in the article \cite{vorontsov2003entropic}, in which
the entropic sampling \cite{lee1993new} was applied within the Wang--Landau algorithm  to calculate the DOS
for a 3D quantum system of harmonic oscillators at a finite temperature. In the path integral formalism quantum
particles are presented as ``ring polymers'' consisting of a lot of ``beads'' connected by harmonic-like bonds (springs) 
\cite{vorontsov2003entropic}. In Ref.~\cite{vorontsov2003entropic} the exact data for the energy and canonical 
distribution were reproduced for a wide range of temperatures.

In this paper we propose a new path integral representation of DOS
in the Wigner formulation of quantum mechanics and the path integral Monte Carlo method (WPIMC) for its calculation. 
We hope that our approach will be a compromise between the accuracy and speed of calculations.
The suggested  approach is applicable for predicting DOS not only for bulk structures (3D)
but also for surfaces (2D) in multi-component systems.
To illustrate the basic ideas we make use of a simple model of strongly coupled soft-sphere fermions that is useful
in statistical mechanics and capable of grasping some physical properties of complex systems. 
This model includes also the one-component plasma (OCP), which is of great astrophysical importance
being an excellent model for describing many features 
of superdense, completely ionized matter \cite{luyten1971review,potekhin2010physics}.
Moreover, theoretical studies of strongly interacting particles obeying the Fermi--Dirac statistics
is a subject of general interest in many fields of physics.  

At strong interparticle interaction perturbative methods cannot be applied, so direct computer simulations have to be used.
At non-zero temperatures the most widespread numerical method in quantum statistics is a Monte--Carlo (MC) method
usually based on the representation of a quantum partition function in the form of path integrals
in the coordinate representation \cite{feynmanquantum,zamalin1977monte}.
A direct computer simulation allows one to calculate the 
thermodynamic properties of dense noble gases, dense hydrogen, electron-hole and quark-gluon plasmas, etc. 
\cite{EbelForFil,ForFilLarEbl,dornheim2018uniform,ceperley1995path,pollock1984simulation,singer1988path,filinov2022solution}.  

The main difficulty of the path integral Monte-Carlo (PIMC) method for Fermi systems is the ``fermionic sign problem''
arising due to the antisymmetrization of a fermion density matrix \cite{feynmanquantum}.
For this reason thermodynamic quantities become small differences of large numbers associated
with even and odd permutations. 
To overcome this issue a lot of approaches have been developed.
In Ref.~\cite{ceperley1991fermion,ceperley1992path}, to avoid the ``fermionic sign problem'', a restricted fixed--node path--integral
Monte Carlo (RPIMC) approach was proposed. In the RPIMC only positive permutations are taken into account, so the accuracy of the results is unknown.
More consistent approaches are the permutation blocking path integral Monte Carlo (PB-PIMC) 
and the configuration path integral Monte Carlo (CPIMC) methods \cite{dornheim2018uniform}. 
In the CPIMC the density matrix is presented as a path integral in the space of occupation numbers.
However it turns out that both methods also exhibit the ``sign problem'' worsening the accuracy of PIMC simulations.
 
In \cite{larkin2017pauli,larkin2017peculiarities} an alternative approach based on the Wigner formulation of 
quantum mechanics in the phase space \cite{wigner1934interaction,Tatr} was used to 
avoid the antisymmetrization of matrix elements and hence the ``sign problem''. This approach allows  
to realize the Pauli blocking of fermions and is able to calculate quantum momentum distribution functions
as well as transport properties \cite{EbelForFil,ForFilLarEbl}. 

Here we propose a path integral representation of DOS in the Wigner phase space. This approach also allows to reduce 
the ``sign problem'' as the exchange interaction is expressed through a positive semidefinite Gram determinant 
\cite{filinov2021monte}. 

In section II we consider the path integral description of quantum DOS in the
Wigner formulation of quantum mechanics. 
In section III we derive a pseudopotential for soft spheres accounting for quantum effects in the interparticle interaction.
In section IV we present the results of our simulations. 
For a 2D quantum system of strongly correlated soft--sphere fermions
we present the DOS, internal energy distributions and spin -- resolved radial distribution functions
obtained by the new path integral Monte Carlo method (WPIMC) (see Supplemental Material)
for the hardness of the soft--sphere potential smaller or of the order of unity.
In section V we summarize the basic results. 


\section{Path integral representation of the density of state}
We consider a 2D system of $N$ soft--sphere particles obeying the Fermi--Dirac statistics.   
The Hamiltonian of the system
${\hat H}={\hat K}+{\hat U}$ contains the kinetic ${\hat K}$ and interaction energy ${\hat U}$ contributions
taken as the sum of pair interactions 
$\phi(r) =\epsilon (\sigma /r)^n$, where $r$ is the interparticle distance, $\sigma$ characterizes the effective particle size,
$\epsilon$ sets the energy scale  and $n$ is a parameter determining the potential hardness. 

Density of state (DOS) is a fundamental function of a system and can be defined as $\Omega(E)=\mbox{Tr}\{\delta(E\hat{\rm I }-\hat{H})\}$ \cite{kubo2012statistical}. $\Omega(E) {\rm d } E $ determines the number of states between $E$ and $E + {\rm d } E$ per unit volume \cite{sese2020real} ($\hat{\rm I }$ is the unit operator, while $\delta$ is the delta function).
The DOS can be used to compute important thermodynamic properties such as, for example,
internal energy, entropy and heat capacity.

In our approach we are going to rewrite $\Omega(E)$ in an identical form using 
the property of the delta function 
\begin{multline}
	\Omega(E)=\mbox{Tr}\{\delta(E\hat{\rm I }-\hat{H})\hat{ \rm I}\}=\mbox{Tr}\{\delta(E\hat{\rm I }-\hat{H})\exp( E\hat{\rm I }-\hat{H}) \} \\
	=\frac{1}{2\pi} \int \hspace{-3pt} {\rm d} \omega \mbox{Tr}\{  \exp \rm i \omega \big(E\hat{ \rm I } -\hat{H} \big) \exp( E\hat{\rm I}- \hat{H}) \}
	=\frac{1}{2\pi} \int \hspace{-3pt} {\rm d} \omega \mbox{Tr}\{  \exp \kappa(\omega) \big(E\hat{\rm I }-\hat{H}\big) \}
	\\
	=\frac{1}{2\pi} \int \hspace{-3pt} {\rm d} \omega  \int \hspace{-3pt} {\rm d} q_1 \left\langle q_1 \left |\exp \kappa(\omega)
	\big(E\hat{\rm I }-\hat{H}\big)\right | q_1 \right\rangle ,
	\label{dt1}  
\end{multline}
where  $\kappa(\omega)=1+\rm i \omega $, angular brackets $\langle q|\tilde{q} \rangle  $ mean the scalar products of the eigenvectors $|q\rangle $
and $|\tilde{q}\rangle $ of the coordinate operator  $\hat{q}$  (  $\langle \hat{q}|q\rangle=q|q\rangle$,
$\langle q|\tilde{q} \rangle=\delta(q-\tilde{q})$), $\hat{\rm I}=\int | q \rangle {\rm d} q \langle q |$ is the unit operator,
$\psi(q)= \langle q|\psi\rangle$ is the wave function \cite{Tatr} ), 
the angular brackets in expression $\langle q_1|A|q\rangle $ mean the scalar products of vectors
$|q_1\rangle$ and  $|\hat{A}|q\rangle$ arising after the action of operator
$\hat{A}$ on vector $|q\rangle$, $\rm i$ is the imaginary unit. Further in the text, it is convenient to imply that energy is expressed in units of $k_B T$
($k_B$ is the Boltzmann constant, $T$ is the temperature of the system) and $q_1$ is a $2N$-dimensional vector of the particle coordinates.

Since the operators of kinetic and potential energy do not commute, an exact explicit
analytical expression for the DOS is unknown but can be constructed using a path integral 
approach~\cite{feynmanquantum,NormanZamalin,zamalin1977monte} based on the operator identity 
$\exp \big( \kappa(\omega) \big(E\hat{\rm I }-\hat{H}\big)\big) = 
\exp \big( \epsilon(\omega) \big(E\hat{\rm I }-\hat{H}\big)\big) \times  \dots \times \exp\big( \epsilon(\omega) \big(E\hat{\rm I }-\hat{H}\big)\big) $ 
with  $\epsilon(\omega)=\kappa(\omega)/M$, where $M$ is a large positive integer. So the DOS can be rewritten in the coordinate representation as 
\begin{multline}
	\Omega(E)=\frac{1}{2\pi} \int \hspace{-3pt} {\rm d} \omega  \int \hspace{-3pt} {\rm d} q_1 
	\left\langle q_1 \left |\exp \kappa(\omega) \big(E\hat{\rm I }-\hat{H}\big)\right | q_1\right\rangle
	=\frac{1}{2\pi} \int \hspace{-3pt} {\rm d} \omega  
	\prod_{j=1}^M \int \hspace{-3pt} {\rm d} q_j {\rm d} \tilde{q}_j \\
	\times 	\left\langle q_1 \left | \exp \frac{\rm i \omega}{M}  \big(E\hat{\rm I }-\hat{H}\big)\right | \tilde{q}_1 \right\rangle
	\left\langle \tilde{q}_1 \left | \exp \frac{1}{M}  \big(E\hat{\rm I }-\hat{H}\big)\right | \tilde{q}_2 \right\rangle 
	\\ \times \left\langle \tilde{q}_2 \left | \exp  \frac{\rm i \omega}{M} \big(E\hat{\rm I }-\hat{H}\big)\right | q_2 \right\rangle  
	\left\langle q_2 \left | \exp \frac{1}{M}  \big(E\hat{\rm I }-\hat{H}\big)\right | q_3 \right\rangle
	\\ \times \left\langle q_3 \left | \exp  \frac{\rm i \omega}{M} \big(E\hat{\rm I }-\hat{H}\big)\right | \tilde{q}_3 \right\rangle  
	\left\langle \tilde{q}_3 \left | \exp \frac{1}{M}  \big(E\hat{\rm I }-\hat{H}\big)\right | \tilde{q}_4 \right\rangle \dots    
	\\ \times  \left\langle \tilde{q}_M \left | \exp \frac{\rm i \omega}{M} \big(E\hat{\rm I }-\hat{H}\big)\right | q_M  \right\rangle 
	\left\langle q_M \left | \exp \frac{1}{M}  \big(E\hat{\rm I }-\hat{H}\big)\right | q_1 \right\rangle , 
	\label{dt2}
\end{multline}
where we have used the coordinate representation of the unit operator $\hat{\rm I}=\int | q \rangle {\rm d} q \langle q |$ \cite{Tatr}.
  
To present the DOS in the Wigner representation of quantum mechanics let us consider
the Weyl symbol of an operator. For example, for the operator $\hat{H}$ the corresponding Weyl symbol is the  Hamiltonian function $ H(pq)$
\cite{wigner1934interaction,Tatr} 
\begin{eqnarray}
	H(pq)=\int \hspace{-3pt} {\rm d} \xi \exp ( {\rm i} \left\langle p |  \xi \right\rangle )  
	\left\langle q-\xi/2 \left |\hat{H} \right | q+\xi/2 \right\rangle,
	\label{dt3}
\end{eqnarray}
where the vectors $\xi$ and momentum $p$ are $2N$--dimensional vectors.

The inverse Fourier transform allows to express matrix elements of operators through their Weyl symbols.
So for large $M$ with the error of the order of $(1/M)^2$ required for the path integral approach \cite{feynmanquantum,zamalin1977monte} we have 
\begin{multline}
	\left\langle Q_j-\xi_j/2  \left | \exp  \frac{\rm i \omega}{M} \big(E\hat{\rm I }-\hat{H}\big)\right | Q_j+\xi_j/2 \right\rangle
	\\
	\approx \left\langle Q_j-\xi_j/2  \left | \hat{\rm I }+ \frac{\rm i \omega}{M} \big(E\hat{\rm I }-\hat{H}\big)\right | Q_j+\xi_j/2 \right\rangle 
	+ {\rm O}\left(\frac{1}{M}\right)^2
	\\
	=\left(\frac{1}{2\pi}\right)^{(2N)} \int \hspace{-3pt} {\rm d} P_j \exp ( - {\rm i} \left\langle P_j |  \xi_j \right\rangle )
	\bigg( 1+ \frac{\rm i \omega}{M} \big(E-H(P_j,Q_j)\big) \bigg)  
	\\
	\approx \left(\frac{1}{2\pi}\right)^{(2N)} \int \hspace{-3pt} {\rm d} P_j \exp\left( - {\rm i} \left\langle P_j |  \xi_j \right\rangle \right)
	\exp \left(\frac{\rm i \omega}{M} \big(E-H(P_j,Q_j)\big) \right)   +  {\rm O}\left(\frac{1}{M}\right)^2,
	\label{dt4}
\end{multline}
where new variable $Q_j$ and $\xi_j$ are defined by equations:    
$Q_j=(\tilde{q}_j + q_j/2)$, $\xi_j=(\tilde{q}_j - q_j)$ for $j=1, \dots, M$  ($q_j=Q_j-\xi_j/2$,  $\tilde{q}_j=Q_j+\xi_j/2$) and  
$H(P_j,Q_j)= \left\langle P_j|P_j \right\rangle/2m + U(Q_j) $   are  the sums of the Hamilton functions for $N$ paricles at a given $j$.
Further for convenience we will use both set of variable ($Q$, $\xi$) and ($q$, $\tilde{q}$).
The final expression for the product is 
\begin{multline}
	\prod_{j=1}^M \left\langle q_{j} \left | \exp  \frac{\rm i \omega}{M} \big(E\hat{\rm I }-\hat{H}\big)\right | \tilde{q}_j \right\rangle 
	\approx \left(\frac{1}{2\pi}\right)^{(2NM)} \prod_{j=1}^{M}  \int \hspace{-3pt}  {\rm d} P_j  \exp ( - {\rm i} \left\langle P_j |  \xi_j \right\rangle )
	\exp \frac{\rm i \omega}{M} \big(E-H(P_j,Q_j)\big).   
	\label{dt5}  
\end{multline} 
Then the DOS is  presented as  
\begin{multline}
	\Omega(E)=\left(\frac{1}{2\pi}\right)^{(2NM)}\frac{1}{2\pi} \int \hspace{-3pt} {\rm d} Q {\rm d} P
	\int \hspace{-3pt} {\rm d} \omega   \exp \big( {\rm i} \omega \big(E-H(P,Q)\big)
	\int \hspace{-3pt} {\rm d} \xi \exp ( -{\rm i} \left\langle P | \xi \right\rangle )
	\\
	\times \left\langle \tilde{q}_1 \left | \exp \frac{1}{M}  \big(E\hat{\rm I }-\hat{H}\big)\right | \tilde{q}_2 \right\rangle 
	\left\langle q_2 \left | \exp \frac{1}{M}  \big(E\hat{\rm I }-\hat{H}\big)\right | q_3 \right\rangle  \cdots
	\\\times
	\left\langle q_M \left | \exp  \frac{1}{M}  \big( E\hat{\rm I }-\hat{H}\big)\right | q_1 \right\rangle,
	\label{dt6}
\end{multline}
where  $H(P,Q)=\sum_{j=1}^{M} H(P_j,Q_j)/M$, 
$Q=\{Q_1 , \dots , Q_{M}\}$ and $P=\{P_1 , \dots , P_{M}\}$, $\xi=\{\xi_1, \dots , \xi_{M}\}$ 
are $2NM$--dimensional vectors and  $\prod_{j=1}^M{\rm d} q_j {\rm d} \tilde{q}_j= {\rm d} Q {\rm d} \xi$.

The final expression for the DOS in the Wigner approach to quantum mechanics  can be written as:
\begin{eqnarray} 
	&&\Omega(E)= \exp (E) \int \hspace{-3pt} {\rm d} Q {\rm d} P 
	\delta (E- H(P,Q) W\left(P,Q\right),
	\label{dt7}
\end{eqnarray}
where $\delta (E - H(P,Q))$  is  the path integral analogue of the  Weyl symbol of the operator
$\delta(E\hat{\rm I }-\hat{H})$ \cite{wigner1934interaction,Tatr}   
\begin{multline}
	\delta (E - H(P,Q))=
	\frac{1}{2\pi} \int \hspace{-3pt} {\rm d} \omega 
	\exp {\rm i} \omega \big(E-H(P,Q)\big) 	\\
	\approx \frac{1}{2\pi} \int \hspace{-3pt} {\rm d} \omega \exp ( \rm i \left\langle P | \xi \right\rangle 
	\prod_{j=1}^M \left\langle q_{j} \left | \exp  \frac{\rm i \omega}{M} \big(E\hat{\rm I }-\hat{H}\big)\right | \tilde{q}_j \right\rangle. 
	\label{dt8}
\end{multline}

So the generalization of the  Wigner function $W\left(P,Q\right) $ is defined as   
\begin{multline}
	W\left(P,Q\right) = (\frac{1}{2\pi})^{(2NM)} \exp (-E)
	\int \hspace{-3pt} {\rm d} \xi \exp ( - \rm i \left\langle P | \xi \right\rangle )  \\
	\times \left\langle \tilde{q}_1 \left | \exp \frac{1}{M}  \big(E\hat{\rm I }-\hat{H}\big)\right | \tilde{q}_2 \right\rangle 
	\left\langle q_2 \left | \exp \frac{1}{M}  \big(E\hat{\rm I }-\hat{H}\big)\right | q_3 \right\rangle  \cdots
	\\\times
	\left\langle q_M \left | \exp  \frac{1}{M}  \big( E\hat{\rm I }-\hat{H}\big)\right | q_1 \right\rangle,
	\label{dt9}
\end{multline}
Herein we have assumed that the operators $\hat{H}$ do not depend on the spin variables. However, the spin variables
$\sigma$  and the Fermi statistics  can be taken into account by the following redefinition of $W\left(P,Q\right)$ in the canonical ensemble
with temperature $T$  
\begin{multline}
	W\left(P,Q\right)  = \frac{1}{Z(\beta)N! \lambda^{2N}  }  \exp(-E)
	\sum_{\sigma}\sum_{\mathcal{P}} (- 1)^{\kappa_{\hat{Pm}}}  {\cal S}(\sigma,  \mathcal{P} \sigma^\prime)	\big|_{\sigma'=\sigma}\, 
	\int \hspace{-3pt} {\rm d} \xi \exp ( - {\rm i} \left\langle P | \xi \right\rangle ) \\
	\times \left\langle \tilde{q}_1 \left | \exp \frac{1}{M}  \big(E\hat{\rm I }-\hat{H}\big)\right | \tilde{q}_2 \right\rangle 
	\left\langle q_2 \left | \exp \frac{1}{M}  \big(E\hat{\rm I }-\hat{H}\big)\right | q_3 \right\rangle  \cdots
	\left\langle q_M \left | \exp  \frac{1}{M}  \big( E\hat{\rm I }-\hat{H}\big)\right | q_1 \right\rangle \\
	= \frac{ 1 }{Z(\beta)N! \lambda^{2N} } \int \hspace{-3pt} {\rm d} \xi \exp ( - {\rm i} \left\langle P | \xi \right\rangle ) \rho^{(1)} \dots \rho^{(M-1)} \\
	\times \sum_{\sigma}\sum_{\mathcal{P}} (-1)^{\kappa_{\mathcal{P}}} 
	{\cal S}(\sigma, \mathcal{P} \sigma^\prime) 	\big|_{\sigma'=\sigma}\,
	\mathcal{P} \rho^{(M)}\big|_{q_{(M+1)}= q_1}
	\label{dt10}
\end{multline}  
where the  sum is taken over all permutations $\mathcal{P}$ with the parity $\kappa_{\mathcal{P}}$,    
index $j$ labels the off--diagonal high--temperature density matrices 
$\rho^{(j)}\equiv \langle Q_j \pm  \xi_j/2  |e^{-\frac{1}{M}{\hat H}}| Q_{(j+1)} \pm \xi_j/2 \rangle$. 
With the error of the order of $1/M^2$ each high--temperature factor can be presented in the form 
$\rho^{(j)}  = \langle Q_j \pm  \xi_j/2  |e^{-\frac{1}{M}{\hat H}}| Q_{(j+1)} \pm  \xi_{(j+1)}/2  \rangle  \approx
e^{-\frac{1}{M}{\hat U(Q_j \pm  \xi_j /2)}}  \rho^{(j)}_0$  
with $  \rho^{(j)}_0=\langle Q_j \pm  \xi_j /2 |e^{-\frac{1}{M}{\hat K}}| Q_{(j+1)} \pm  \xi_{}(j+1) /2 \rangle $, 
arising from neglecting the commutator $\left[K,U\right] / (2M^2)$ and further terms of the expansion.
In the limit $M\rightarrow \infty$ the error of the whole product of high temperature factors 
tends to zero $(\propto 1/M)$
and we have an exact path integral representation of the Wigner functions. 

The partition function $Z$ for a given temperature $T$ and volume $V$ can be similarly expressed as 
\begin{equation}\label{q-def}
	Z(\beta) = \frac{1}{N! \lambda^{2N} } \sum_{\sigma}\int\limits_V {{\rm d}} Q_1 \,\rho(Q_1 , \sigma;\beta),
\end{equation}
where $\rho(Q_1, \sigma;\beta)$ denotes the diagonal matrix
elements of the density operator ${\hat \rho} = e^{-{\hat H}}$ and 
$\lambda=\sqrt{\frac{2\pi\hbar^2\beta}{m}}$ is the thermal wavelength and $\beta=1/k_BT$.
The integral in Eq.~(\ref{q-def}) can be rewritten as 
\begin{multline}
		\sum_{\sigma} \int\limits {{\rm d}} Q_1\,
		\rho(Q_1,\sigma;\beta) 
		\\
		=\int\limits  {{\rm d}} Q_1 \dots {{\rm d}} Q_M \, \rho^{(1)} \dots \rho^{(M-1)}
		\sum_{\sigma}\sum_{\mathcal{P}} (-1)^{\kappa_{\mathcal{P}}} 
		{\cal S}(\sigma, \mathcal{P} \sigma^\prime) 
		\mathcal{P} \rho^{(M)}\big|_{Q_{(M+1)}= Q_1, \sigma'=\sigma}\,  
		\\\approx {} \int\limits  {{\rm d}} Q_1 \dots {{\rm d}} Q_M \, 
		\exp\Biggl\{-\sum_{j=1}^{M-1}\biggl[ 
		\pi \left|Q_j -Q_{(j+1)} \right|^2
		+ \frac{1}{M}U(Q_j)  \biggr]\Biggr\} 
		\mathrm{det} \|\Psi(Q_{M},Q_{1}\|,
		\label{rho-pimc}
\end{multline} 
where we imply that momentum and coordinate are dimensionless variables
$ \tilde{p\lambda}/ \hbar$ and $q/ \tilde{\lambda}$ related to a temperature $MT$  
($\tilde{\lambda}=\sqrt{2\pi\hbar\beta / (m M)}$). 
Spin gives rise to the standard spin part of the density matrix 
${\cal S}(\sigma, \mathcal{P} \sigma^\prime)=\prod_{k=1}^N \delta(\sigma_k,\sigma_{\mathcal{P}k})$, 
($\delta(\sigma_k,\sigma_t)$ is the Kronecker symbol)  with exchange effects accounted for by the permutation
operator  $\mathcal{P}$  acting on coordinates of particles
$\tilde{q}_{(M+1)} $ and spin projections $\sigma'$.      

In the thermodynamic limit the main contribution in the sum over spin variables comes from the term related
to the equal numbers ($N/2$) of fermions with the same spin projection  \cite{EbelForFil,ForFilLarEbl}. 
The sum over permutations gives  the product of determinants 
$ \mathrm{det} \|\Psi(Q_{M},Q_{1}\| = \mathrm{det} \bigl\|e^{-{\pi} \left|Q_{M}^{(k)}-Q_{1}^{t}\right|^2}\bigr\|_1^{N/2} 
\mathrm{det} \bigl\|e^{-{\pi} \left|Q_{M}^{(k)}-Q_{1}^{t}\right|^2}  \bigr\|_{(N/2+1)}^{N}$.   


In general the complex-valued integral over $\xi$ in the definition of the Wigner function (\ref{dt10})
can not be calculated analytically and is inconvenient for Monte Carlo simulations. 
The second disadvantage is that Eqs.~(\ref{dt10}), (\ref{rho-pimc})
contain the sign--altering  determinant $ \mathrm{det} \|\Psi(Q_{M},Q_{1}\|$, which is the reason of the  
``sign problem'' worsening the accuracy of PIMC simulations.
To overcome these problems let us replace the variables of integration 
$Q_j$ by $\zeta_{j} $ 
for any given permutation $\mathcal{P}$ using the substitution
\cite{LarkinFilinovCPP,larkin2017peculiarities}  
\begin{eqnarray}
	Q_j = ( \tilde{P} Q_{(M+1)} -Q_1)\frac{j-1}{M}+Q_1 + \zeta_{j} ,  
	\label{var}
\end{eqnarray}
where $\tilde{P}$ is the matrix representing the operator of permutation $\mathcal{P}$ and equal to the unit matrix $E$ with appropriately transposed columns.
This replacement presents each trajectory $Q_j$  as the sum of the ``straight line'' 
$( \tilde{P} Q_{(M+1)} -Q_1)\frac{j-1}{M}+Q_1 $   
and the deviation $\zeta_j $ from it (here we assume $Q_{(M+1)}= Q_1 $, $j=1, \cdots, M+1$).   

As a consequence the matrix elements of the density matrix can be rewritten  in the form of a path integral 
over \emph{``closed''}  trajectories $ \{  \zeta_{1}, \dots, \zeta_{{(M+1)}} \}$ with $\zeta_{1} =\zeta_{(M+!)} =0 $ (`ring polymers''). 
By making use the approximation for potential $U$ arising from the Taylor expansion up to the first order in the $\xi$,  
and after the integration over $\xi$ \cite{LarkinFilinovCPP,larkin2017peculiarities} 
and some additional transformations (see  \cite{filinov2022bound,filinov2020uniform,filinov2021monte} for details)  
the Wigner function can be written in the form containing the  Maxwell distribution with quantum corrections 

\begin{multline}
		W(P,Q) \approx \frac{\tilde C(M)}{Z(\beta)N!}
        \exp\Bigl[ - \sum\limits_{j = 1}^{M}  
		\biggl( \pi | \eta_j |^2 + \frac{1}{M} U \biggl(Q_1+\zeta_j\biggr)\biggr) \Bigr]
		\\
		\times \exp\Biggl\{\frac{M}{4 \pi}\sum\limits_{j = 1}^{M} 
		\left< {\rm i } P_j + (-1)^{(j-1)}\frac{1}{2M} 
		\frac{\partial  U (Q_1+\zeta_j ) }{\partial Q_1} 
		\Bigg|  {\rm i } P_j + (-1)^{(j-1)}\frac{1}{2M} 
		\frac{\partial  U (Q_1+\zeta_j ) }{\partial Q_1}	\right>
		\Biggr\}
		\\
		\times \mathrm{det} \|\tilde{\phi}_{kt} \bigl\|_1^{N/2} \mathrm{det}\bigr\|\tilde{\phi}_{kt} \|_{(N/2+1)}^{N_e} , \,
		\label{rho-pimc44} 
\end{multline}
where 
\begin{equation} 
	\tilde{\phi}_{kt} =  \exp \{-{\pi} \left|r_{kt}\right|^2/M \}   
	\exp \Biggl\{-\frac{1}{2M}\sum\limits_{j = 1}^{M}
	\biggl( \phi \Bigl(\Bigl|r_{tk}\frac{2j}{M} +r_{kt} + (\zeta^k_{j}-\zeta^t_{j})\Bigr|  \Bigr) 
	- {} \phi \Bigl(\Bigl|r_{kt} + (\zeta^k_{j}-\zeta^t_{j})\Bigr| \Bigr) \biggr)\Biggr\}, 
	\label{rho-pimcc} 
	\nonumber
\end{equation}
$\eta_j \equiv  \zeta_j  - \zeta_{(j+1)} $, 
$r_{kt}\equiv (Q_1^k-Q_1^t)$, $(k,t=1,\dots,N)$.   
The constant $\tilde{C}(M)$ is canceled in Monte Carlo calculations. 

Let us stress that approximation (\ref{rho-pimc44}) have the correct limits to the cases of weakly and strongly degenerate
fermionic systems. Indeed, in the classical limit the main contribution comes from the diagonal matrix elements due to the factor
$\exp \{-{\pi} \left|r_{kt}\right|^2/M \}$
and the differences of potential energies in the exponents
are equal to zero (identical permutation).  
At the same time, when the thermal wavelength is of the order of the average interparticle distance 
and the trajectories are highly entangled the term $r_{tk}\frac{2j}{M}$ (breaking `ring polymers'') in the potential energy 
$ \phi \Bigl(\Bigl|r_{tk}\frac{2j}{M}  +r_{kt} + (\zeta^k_{j}-\zeta^t_{j})\Bigr|  \Bigr)$ can be neglected 
and the differences of potential energies in the exponents tend to zero \cite{filinov2020uniform, filinov2021monte}. 

Thus, the problem is reduced to calculating the matrix elements of the density matrix 
$\rho=\exp{(-\beta\hat{H}})$, which is similar to the simulation of thermodynamic properties and, according to Eq.~(\ref{dt7}),
the problem of DOS calculation is reduced to considering the internal--energy histogram in the canonical ensemble multiplied by $\exp (E)$.
 
\section{Quantum pseudopotential for soft--sphere fermions}\label{PsdPtt}
The high--temperature density matrix $\rho^{(j)}=\langle
r^{(j)}|e^{-\epsilon {\hat H}}|r^{(j+1)}\rangle$ can be expressed as a
product of two--particle density matrices \cite{EbelForFil} 
\begin{equation} 
	\label{rho_ab}
	\rho(r_{l},r'_{l}, r_{t}, r'_{t};\epsilon)
	= \frac{1}{\tilde{\lambda}^6}
	\exp\left[-\frac{\pi}{\tilde{\lambda}^2} |r_{l} - r'_{l}|^2\right] \\
\exp\left[-\frac{\pi}{\tilde{\lambda}^2} |r_{t} - r'_{t} |^2\right]
	\exp[-\epsilon \Phi^{OD}_{lt}]\,.
\end{equation} 
This formula  results from the factorization of the density matrix into the kinetic and potential parts,
$\rho \approx\rho_0^K\rho^{U}$. The off--diagonal density matrix element (\ref{rho_ab}) involves
an effective pair interaction by a pseudopotential, which can be expressed approximately via
the diagonal elements, $\Phi^{OD}_{lt}(r_{l},r'_{l},r_{t},r'_{t};\epsilon)
\approx [\Phi_{lt}(r_{l}-r_{t};\epsilon)+\Phi_{lt}(r'_{l}-r'_{t};\epsilon)]/2$. 
\begin{figure}[htp]
	\centering
	\includegraphics[width=0.6\columnwidth,clip=true]{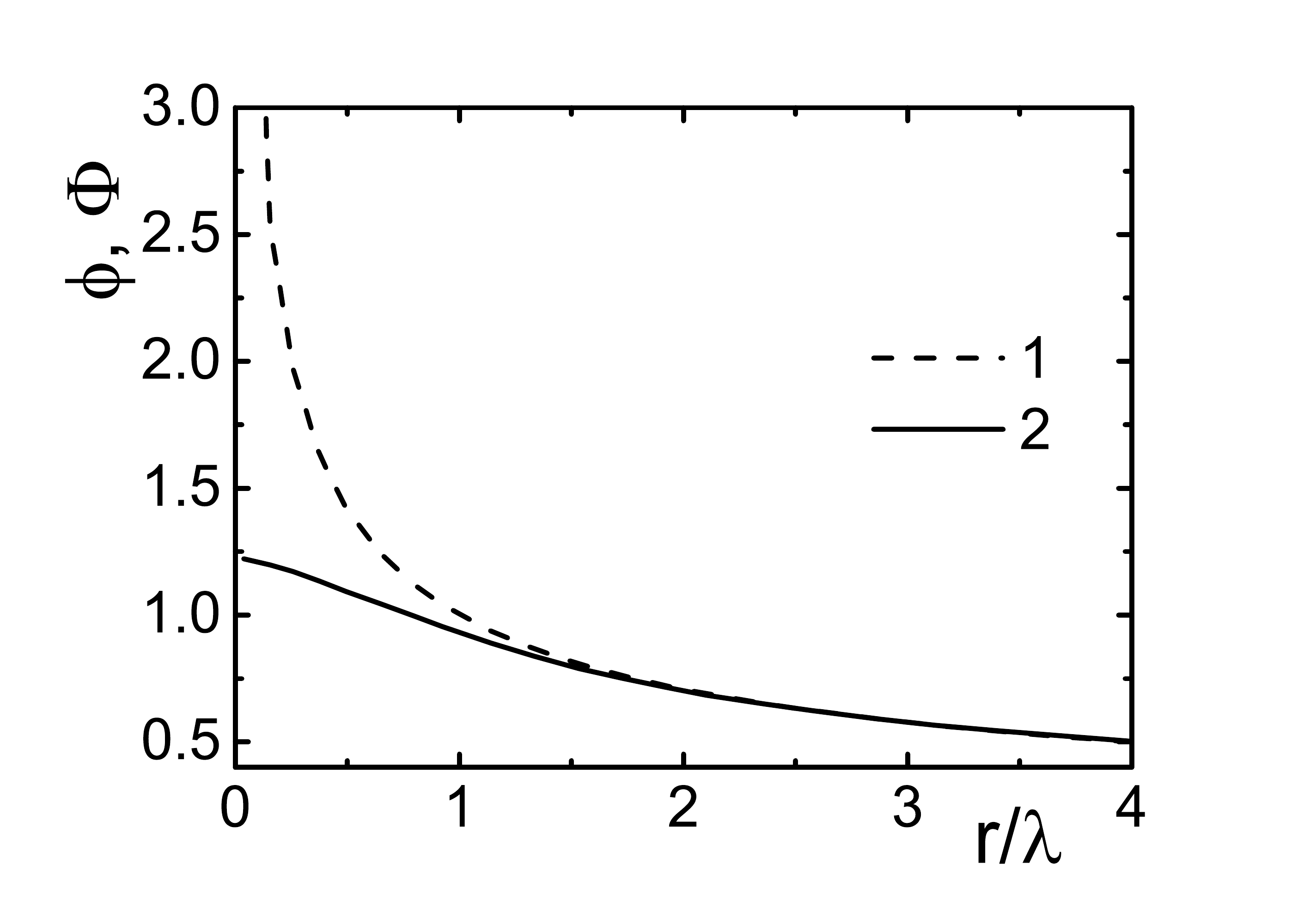}
	\caption{(Color online) The soft--sphere potential $\phi$ (line 1) at $n=1/2$ and respective pseudopotential $\Phi$ (line 2)
		defined by Eq.~(\ref{kelbg-d}) in conditional units.
		\label{avrpt} 
	}
\end{figure} 

To estimate $\Phi(r)$ for each high-temperature density matrix we use the Kelbg functional \cite{demyanov2022derivation,Ke63}
allowing to take into account quantum effects in interparticle interaction.
The peudopotential $\Phi(r)$  is defined by the Fourier transform $v(t)$ of the potential $\phi(r)$. 
This transform can be found at $n < 3$ for the corresponding Yukawa--like potential $\exp(-\kappa r)/r^n$
in the limit of ``zero screening''   ( $\kappa \rightarrow 0$ ) 
\begin{equation}
	v(t) = \frac{4 \pi t^n \Gamma(2-n)\sin (n\pi/2 )}{t^3}, 
\end{equation}
where $\Gamma$ is the gamma function. The resulting quantum pseudopotential has the following form 
\begin{equation} 
	\Phi(r) =  \frac{\sqrt{\pi}}{8\pi^3} 
	\int^{\infty}_0 v(t) \exp (-(\tilde{\lambda} t)^2/4) 
	\frac{ \sin (t r) {\rm erfi}(\tilde{\lambda} t/2)}{(t r) \tilde{\lambda}t} 4 \pi t^2 {\rm d}t,
	\label{kelbg-d}
\end{equation} 
where ${\rm erfi(z)}= \rm i \, efr(\rm i z)$, ${\rm erf(z)}$  is the error function \cite{demyanov2022derivation}. 
This pseudopotential is finite at zero interparticle distance, $\Phi(0)=\lambda^{-n}\Gamma(1-n/2)$,  and decreases
according to the power law $(\lambda /r)^n$ for distances larger than the thermal wavelength (see Figure~\ref{avrpt}).

For more accurate accounting for quantum effects the ``potential energy''  $U(q^{(j)},q^{(j+1)})$ 
in (\ref{dt10}) and (\ref{rho-pimc}) has to be taken as the sum of pair interactions given  
by $\Phi^{OD}$ with $\Phi(r)$. 
The pseudopotential $\Phi$ was also used in the Hamilton function $H(pq)$
in the Weyl's symbol of the operator $\delta(E\hat{\rm I }-\hat{H})$.    

However, if the effective hardness of the pseudopotential $\Phi$ is less than $3$  the corresponding energy $\sum_{m=0}^{M-1} \epsilon U(x^{(j)})$
may be divergent in the thermodynamic limit.    
To overcome this deficiency let us modify the pseudopotential  $\Phi$
according to the transformation considered in \cite{hansen1973statistical}  
\begin{equation} 
	\tilde{\Phi}(r) =   [ \Phi(r) - 
	\frac{1}{V}\int\limits_{V}{\rm d^3} y \Phi (r+y) ] \\
	{} =\int\limits {\rm d^3} y \Phi(r+y) \left(\delta (y) - \frac{1}{V}\right).
	\label{kelbg-d2}
\end{equation} 
Here the uniformly ``charged'' background is introduced to compensate the possible divergence of $U(q^{(j)})$
similar to the case of one-component Coulomb plasma.

Let us note that the pseudopotential $\Phi$ corresponding to the Coulomb potential with hardness $n=1$ was often used in PIMC simulations 
of one-- and two--component plasma media in \cite{Ke63,kelbg,afilinov-etal.04pre,ebeling_sccs05,KTR94,EbelForFil,ForFilLarEbl} 
showing good agreement with the data available in the literature.
  

\section{Results of simulations}\label{simulations}  

In this section we investigate the dependence of a radial distribution function (RDF) and DOS on the hardness of the soft--sphere potential. Here, as an interesting example, we present RDFs,
internal--energy distributions (histograms) and DOS for the 2D system of Fermi particles strongly interacting
via the soft sphere potential with hardness $n$ equal to 0.2, 0.6, 1, and 1.4.
The density of soft spheres is characterized by the parameter $r_s=a/\sigma$, defined as the ratio
of the mean distance between the particles $a=\left[1/(\pi \tilde{\rho}) \right]^{1/2}$ to $\sigma$  
($\tilde{\rho}$ is the 2D particle density, $r_s=1/\sqrt{\pi n\sigma^2}$ is the 2D Brucker parameters).
For example, the results presented below have been obtained for the following physical parameters
used in \cite{filinov2022solution} for PIMC simulations of helium-3:
$\epsilon=26.7 {\rm K} $, $\sigma = 5.19 \, a_{\rm B}$ ($a_{\rm B}$ is the Bohr radius),
$m_a = 3.016$  (the soft--sphere mass in atomic units) and $r_s$ is of the order of $2$.

The RDF \cite{kirkwood1935statistical,fisher1964statistical},
internal--energy distribution functions (IED) and DOS can be written in the form  
\begin{eqnarray}  \label{gab-rho}
	&&g_{ab}(r) = \int {\rm d} P {\rm d}Q \,\delta(|q_{1,a}-q_{1,b}|-r)\,
	W(P,Q),
	\nonumber\\ &&
    W(E)=\int \hspace{-3pt} {\rm d} P {\rm d}Q   \, \delta (E- H(P,Q) W\left(P,Q\right),
	\nonumber\\ &&
    \Omega(E)= \exp(E) W(E) ,\,
 \end{eqnarray}
where $E$ and $H(P,Q)$ are energy per particle, $a$ and $b$ label the spin value of a fermion.
The RDF $g_{ab}$ is proportional to the probability density to find a pair of particles of types $a$ and $b$
at a certain distance $r$ from each other. In an isotropic system the RDF depends only on the difference of coordinates because of the
translational invariance.
The DOS gives the number of states in the phase space per an infinitesimal range of internal energy.
In a non-interacting 2D system $\Omega=const$ \cite{kubo2012statistical} and  $g_{ab}  \equiv 1 $, whereas interaction
and quantum statistics result in the redistribution of particles and non-constant DOS.

\begin{figure}[htp] 
	\centering
	\includegraphics[width=0.45\columnwidth,clip=true]{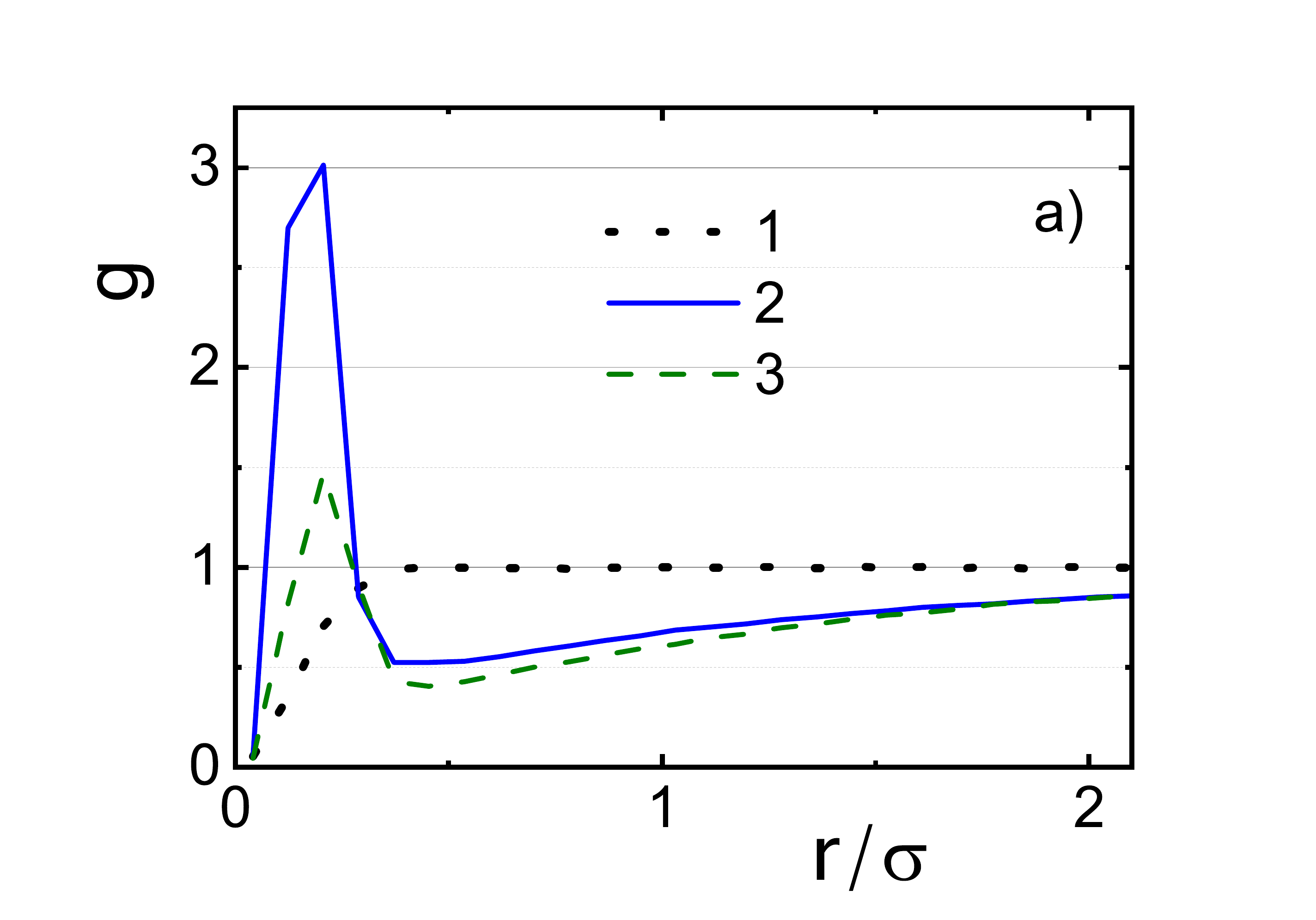}	
	\includegraphics[width=0.45\columnwidth,clip=true]{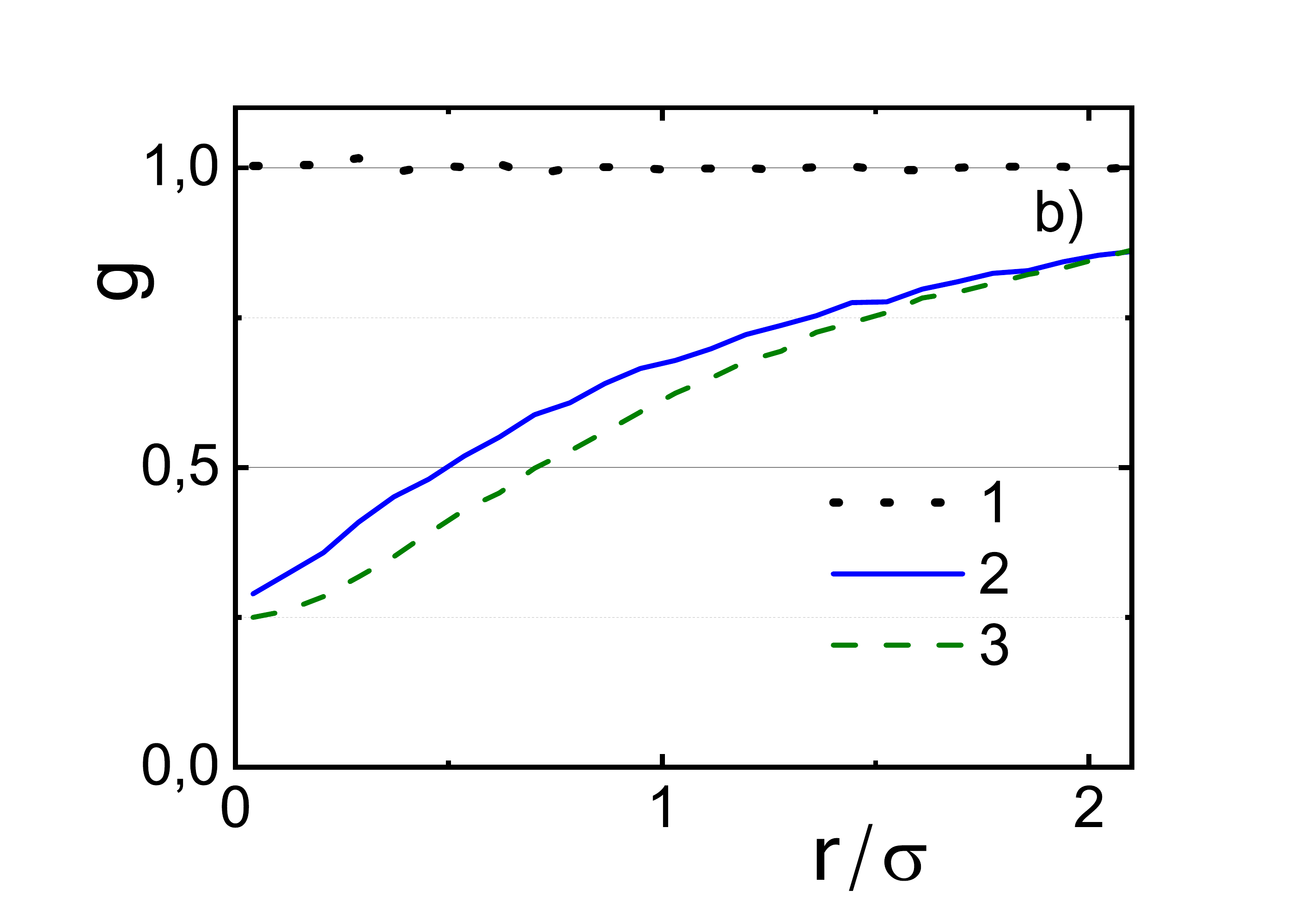}		
	\caption{(Color online) 
		The RDFs for the system of soft--sphere fermions at a fixed density $r_s=2.2$ and temperature $T=60$~K. \\
		Panel a) --- fermions with the same spin projections,
		panel b) --- fermions with the opposite spin projections.
		Lines: 1---ideal system;  2---$n=0.6$; 3---$n=1$. 
		Small oscillations indicate the Monte-Carlo statistical error.
		\label{gdos1} %
	}
\end{figure} 

\begin{figure}[htp] 
	\centering
	\includegraphics[width=0.45\columnwidth,clip=true]{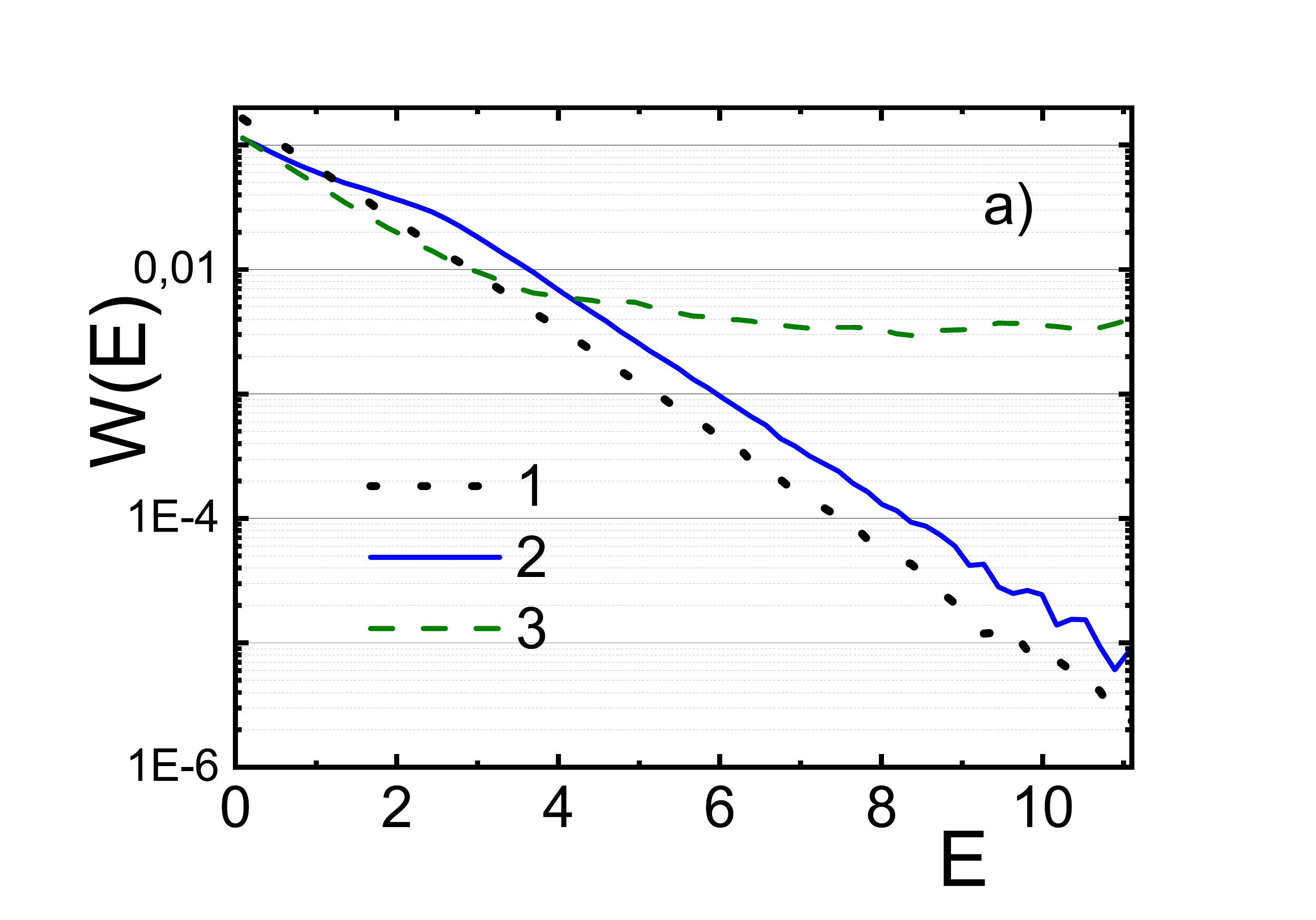}		
	\includegraphics[width=0.45\columnwidth,clip=true]{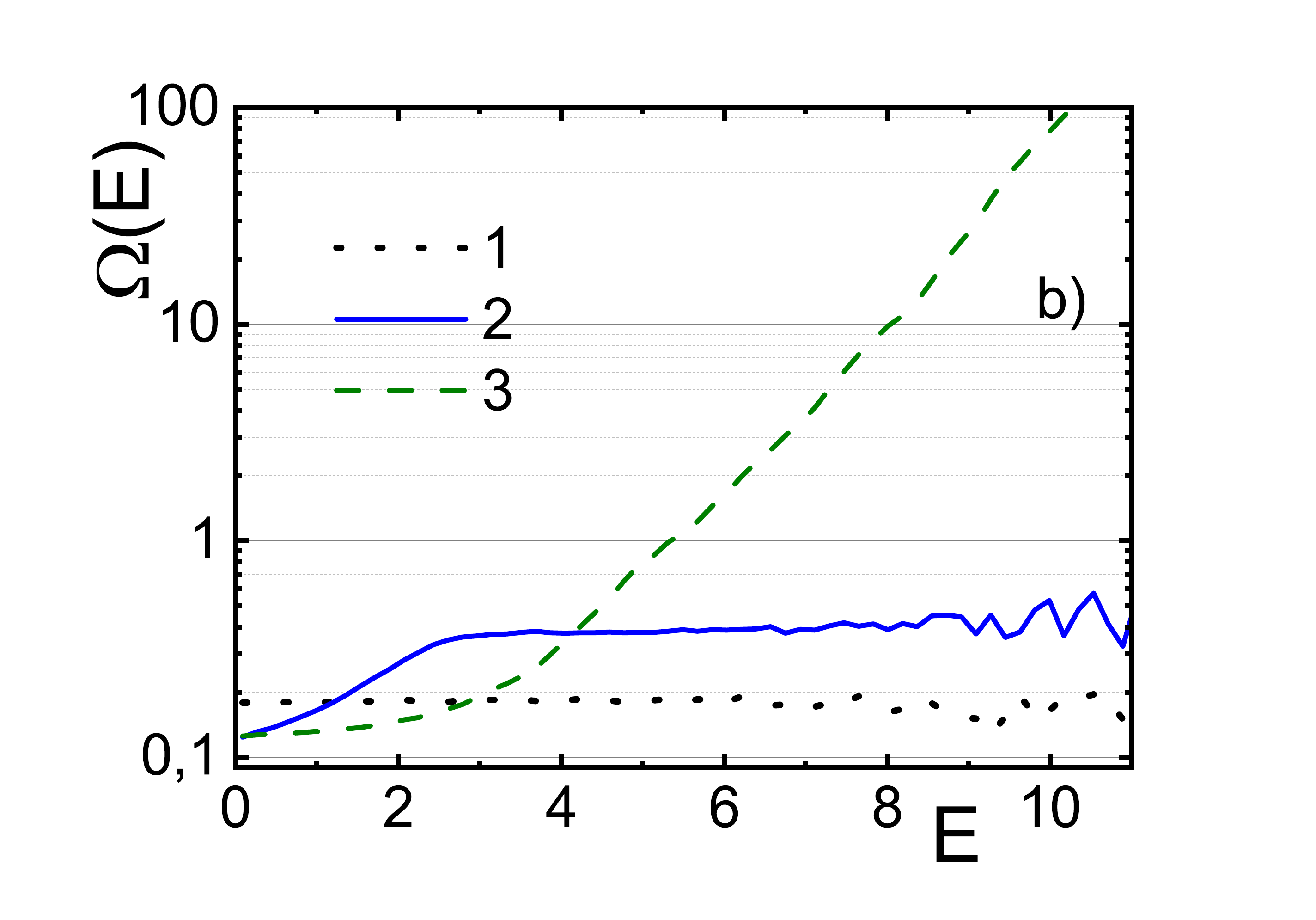}	
	\caption{(Color online) 
		The energy distribution $W(E)$ (panel a) and DOS (panel b) for the system of soft-sphere fermions at a fixed density
		$r_s=2.2$ and temperature $T=60$~K ($\int W(E){\rm d} E=1$,  $\Omega(E)$ in conditional units). Lines: 1---ideal system;  2---$n=0.6$; 3---$n=1$. 
		Small oscillations indicate the Monte-Carlo statistical error.	
		\label{gdos22} %
	}
\end{figure}

Figure~\ref{gdos1} presents the results of our WPIMC calculations for the spin--resolved RDFs
for a fixed density and temperature but at different hardnesses of the soft--sphere potential.
Let us discuss the difference revealed between the RDFs with the same and opposite spin projections.
At small interparticle distances all RDFs tend to zero due to the repulsion nature of the soft--sphere potential.
Additional contribution to the repulsion of fermions with the same spin projection at distances of the order of
the thermal wavelength (lines 1,2,3) is caused by the Fermi statistics effect described by the exchange determinant in (\ref{rho-pimc44})
that accounts for the interference effects of the exchange and interparticle interactions.
This additional repulsion leads to the formation of cavities (usually called exchange--correlation holes)
for fermions with the same spin projection and results in the formation of high peaks on the corresponding RDFs
due to the strong excluded volume effect \cite{barker1972theories}.
The RDFs for fermions with the same spin projection show that the characteristic ``size'' of an exchange--correlation cavity 
with corresponding peaks is of the order of the soft--sphere thermal wavelength ($\lambda/\sigma\sim 0.5$)
that is here less than the average interparticle distance $r_s=2.2$.
Let us stress that the strong excluded volume effect was also observed in the classical systems of repulsive particles
(system of hard spheres) seventy years ago in \cite{kirkwood1950radial} and was derived analytically for 1D case in \cite{fisher1964statistical}.

Note that for fermions with the opposite spin projections the interparticle interaction is not enough to form any peaks
on the RDF.
At large interparticle distances the RDFs decay monotonically to unity due to the short--range repulsion of the potential.

Panels a) and b) of Figure~\ref{gdos22} present the results of WPIMC calculations for the
internal--energy distributions $W(E)$
and DOS $\Omega(E)$
for the same parameters as in Figure~\ref{gdos1}.
Here all $W(E)$ are normalized to unity
($\int \hspace{-3pt} {\rm d} E  \, W(E) = 1 $).

In an ideal system the internal energy of chaotic particle configurations is defined by the Maxwell distribution.
Soft--sphere repulsive interaction increases the energy of any given phase space configuration in comparison with the same
configuration of the ideal system. As the energy distribution is proportional to the fraction of phase space
states with the energy equal to $E$, then this fraction ($W(E)$) have to be shifted to a greater energy,
which we can see in panel a) of Fig.~\ref{gdos22}. The characteristic value of this energy shift is of the order of the product  of
characteristic values of the RDF and pseudopotential at small interparticle distances as the main
contribution to the shift is defined by the region where $g(r)-1$ is nonzero.
Let us remind that both $\Phi(r)$ and $\Phi(0)$ increase  with the hardness
for small interparticle distances ( $\Phi(0)=\lambda^{-n}\Gamma(1-n/2)$,
$\Gamma$ is gamma function), while the RDFs according to panel a) of Fig.~\ref{gdos1}
decrease. This effect is the physical reason of the nontrivial behavior of the DOS in panel b) of Fig.~\ref{gdos22}
 (lines 2, 3) in comparison with the DOS of ideal system, which is identically equal to a constant (line 1) \cite{kubo2012statistical}.

\begin{figure}[htp]
	\centering
	\includegraphics[width=0.45\columnwidth,clip=true]{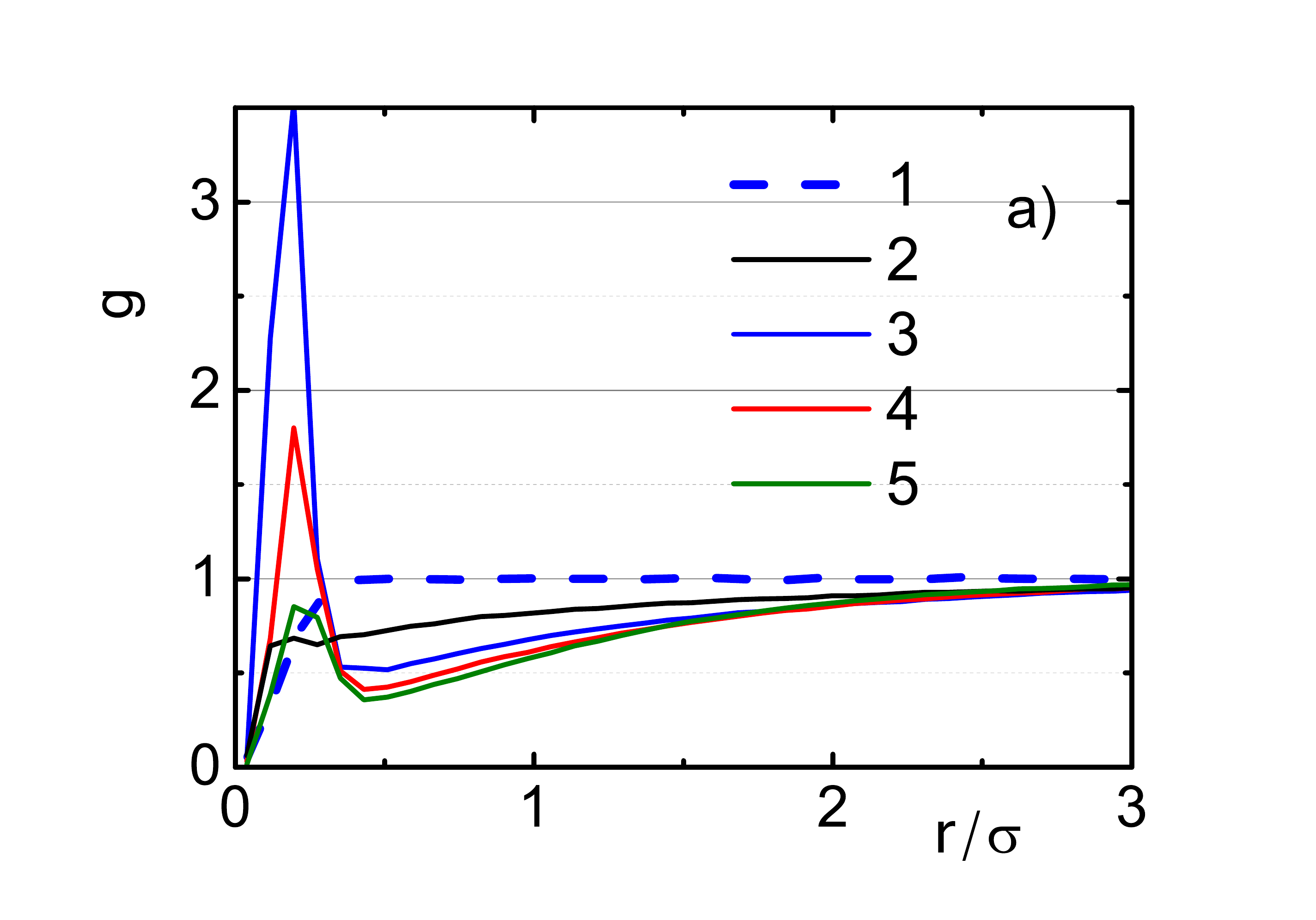}		
	\includegraphics[width=0.45\columnwidth,clip=true]{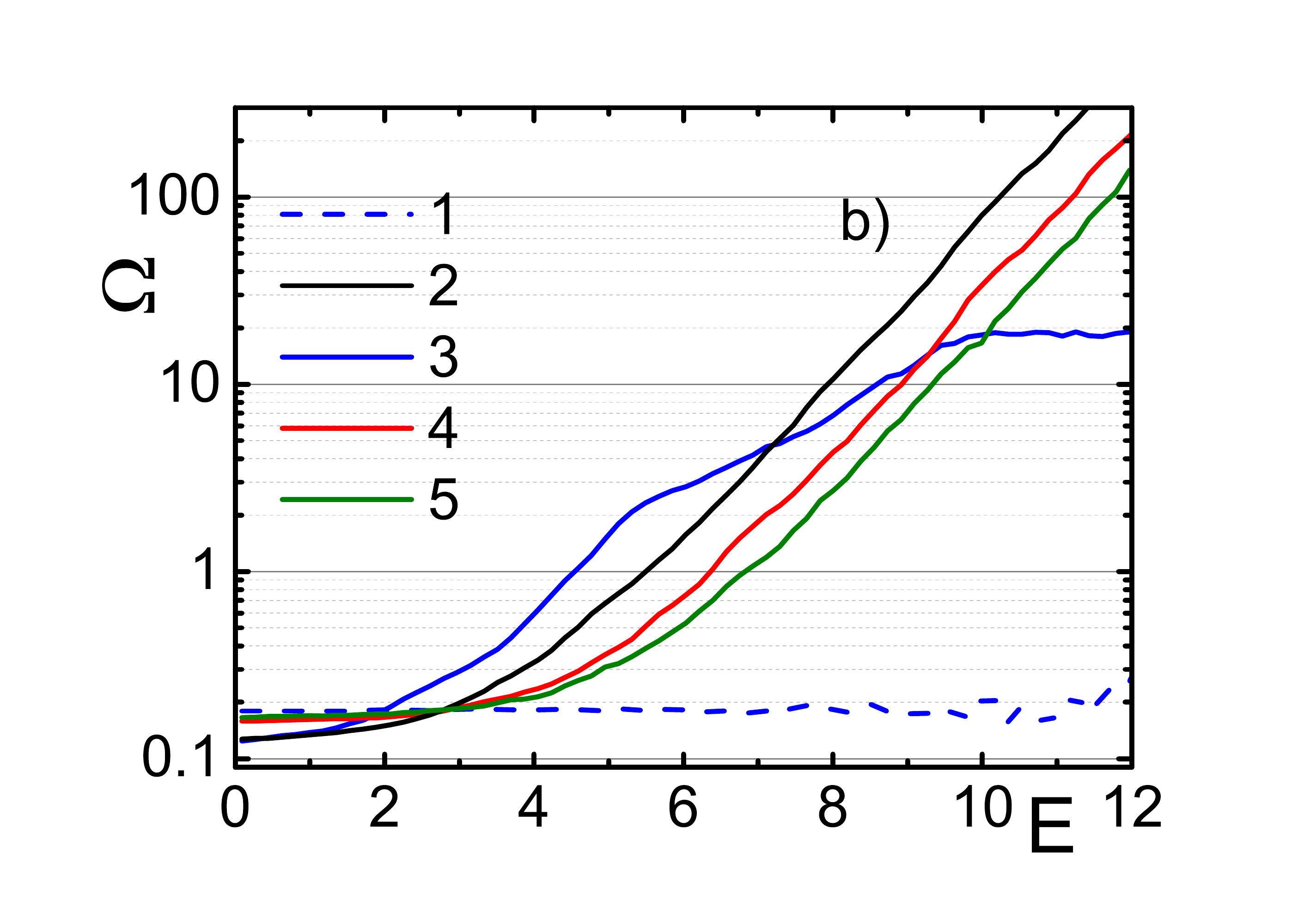}	
  	\caption{(Color online)
The RDFs for the same spin projections (panel a) and DOS (panel b ) for a system of soft-sphere fermions  at a fixed density
 $r_s=2.1$ and temperature  $T=60$~K. \\ Lines: 1---ideal system; 2 ---$n=0.2$; 3---$n=0.6$; 4---$n=1.0$; 5---$n=1.4$.  \\
Small oscillations indicate the Monte-Carlo statistical error.
  		\label{gdos2}
	}
\end{figure}

Panel a) in Fig.~\ref{gdos2} shows the RDFs at the same temperature but for a slightly higher density
($r_s=2.1$ instead of  $r_s=2.2$) and an extended range of hardness in comparison with panel b) of Fig.~\ref{gdos22}.
As in the previous case the heights of the RDF peaks demonstrate non-monotonic behavior and reach the maximum values at $n=0.6$,
while the widths of the peaks are practically the same being of the order of the fermion thermal wavelength.
Generally speaking the changes in the DOS behavior are non-monotonic and show interesting peculiarities at $n=0.6$.  

\begin{figure}[htp] 
	\centering
	\includegraphics[width=0.45\columnwidth,clip=true]{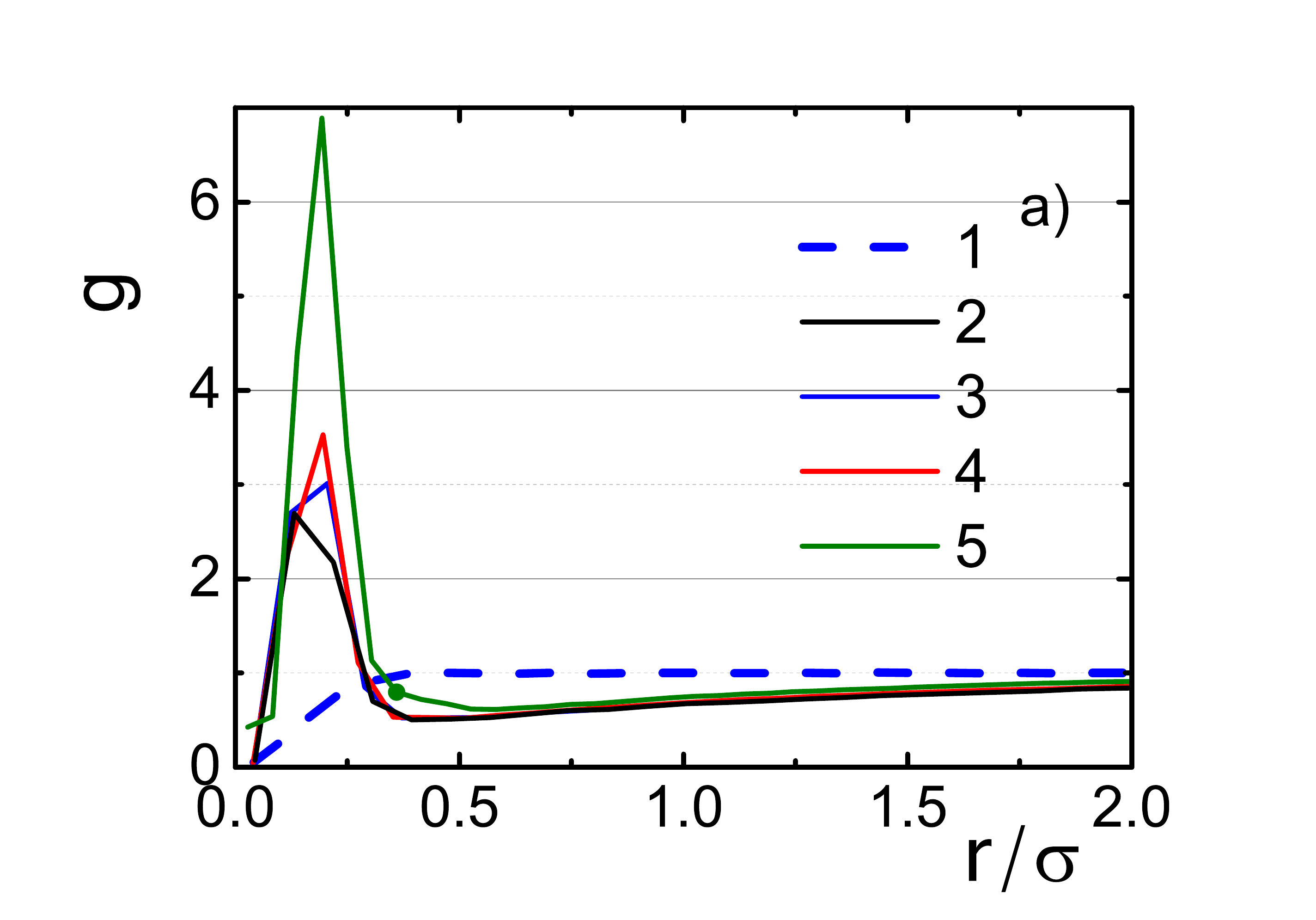}		
	\includegraphics[width=0.45\columnwidth,clip=true]{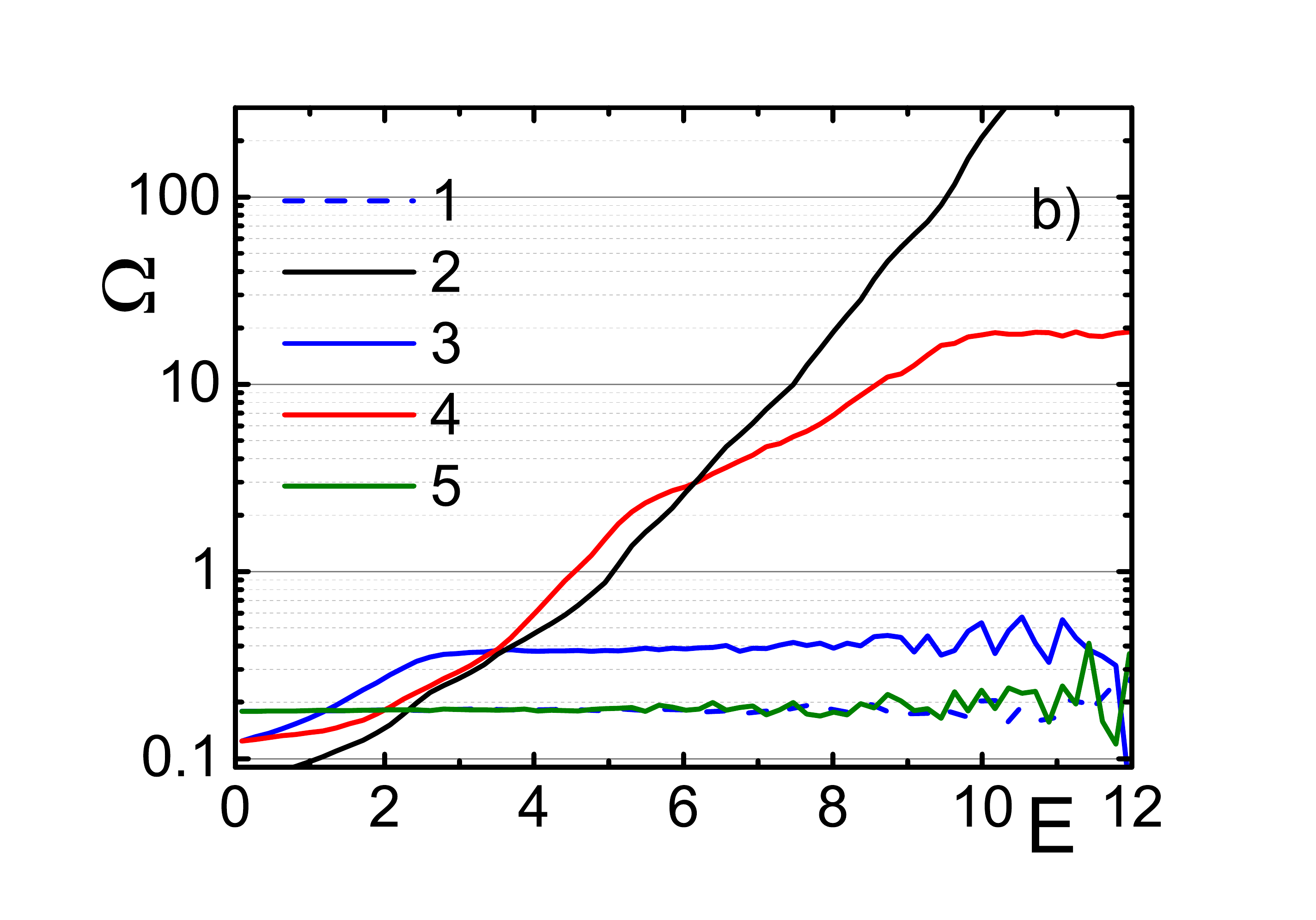}	
 	\caption{(Color online)
The RDFs for the same spin projections (panel a) and $\Omega(E)$ (panel b ) for the system of ideal and interacting
soft-sphere fermions at $n=0.6$, $T=60$~K and different densities $r_s$. \\
Lines: 1---ideal system; 2---$r_s=2.3$; 3---$r_s=2.2$; 4---$r_s=2.1$; 5---$r_s=1.47$.  \\
Small oscillations indicate the Monte-Carlo statistical error.
 		\label{gdos3}
	}
\end{figure}

\begin{figure}[htp] 
	\includegraphics[width=0.45\columnwidth,clip=true]{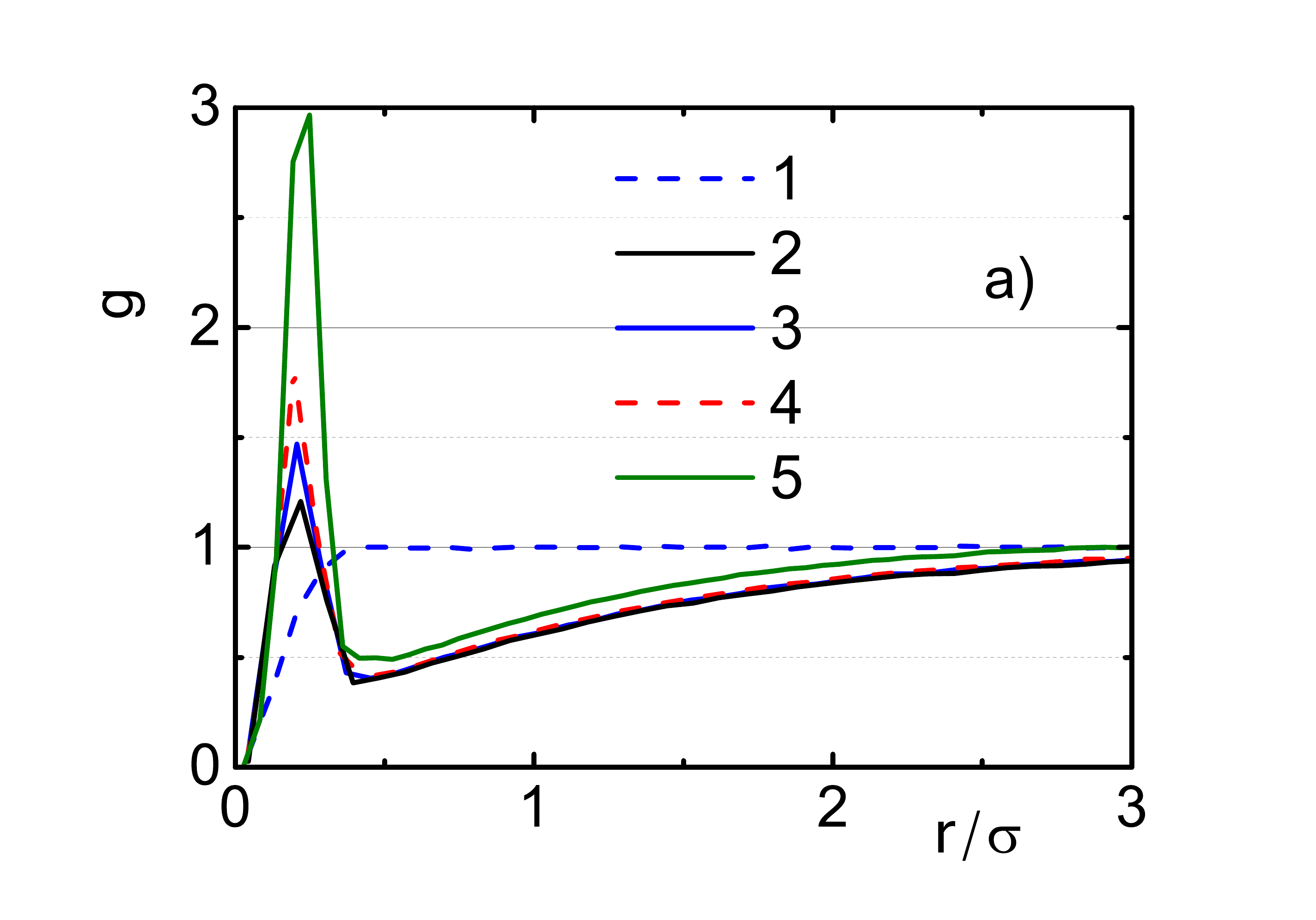}		
	\includegraphics[width=0.45\columnwidth,clip=true]{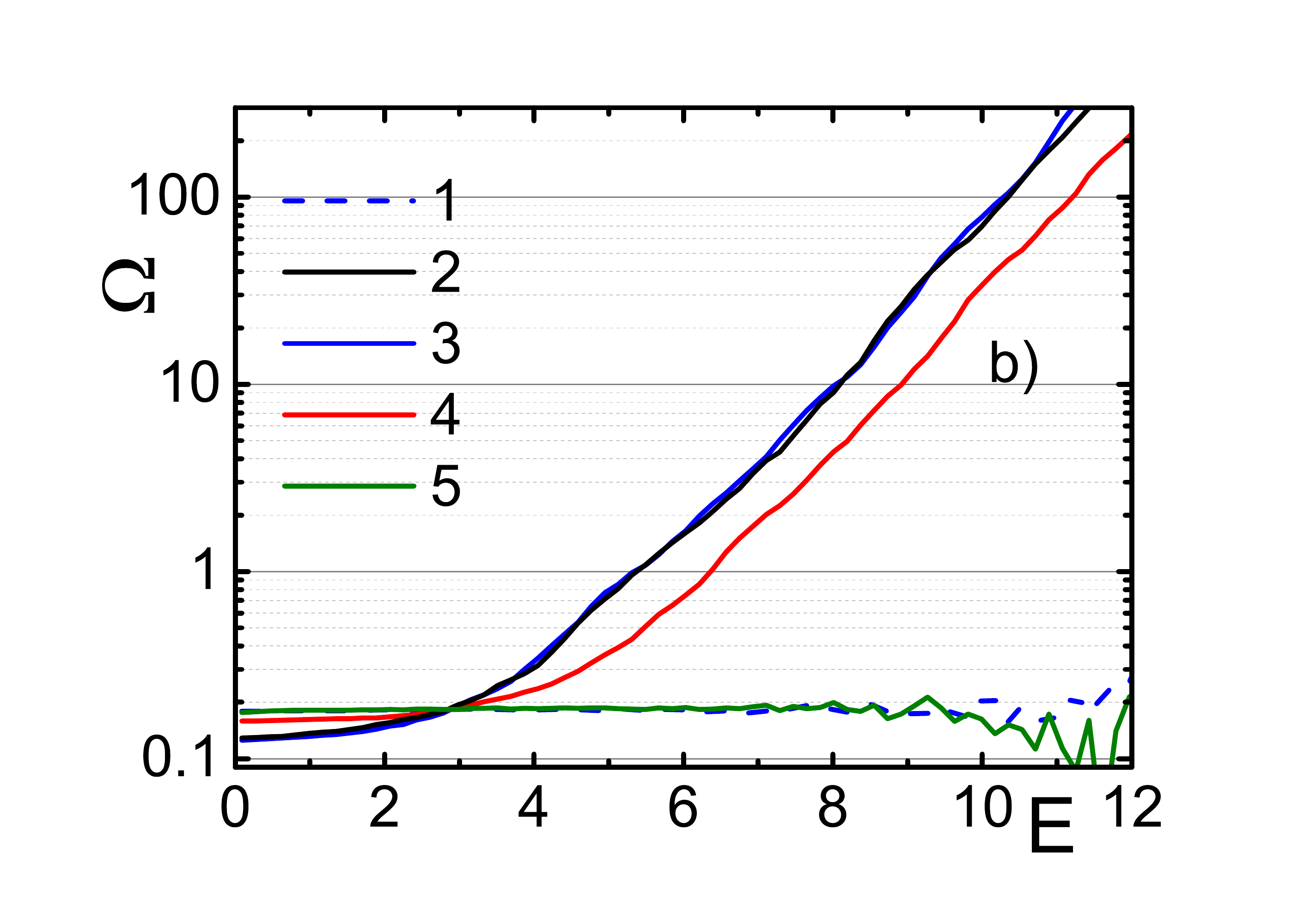}	
 	\caption{(Color online)
The RDFs for the same spin projections (panel a) and DOS (panel b ) for a system of ideal and interacting
soft-sphere fermions at $n=1$, $T=60$~K and different densities $r_s$. 
Lines: 1---ideal system; 2---$r_s=2.3$; 3---$r_s=2.2$; 4---$r_s=2.1$; 5---$r_s=1.47$. 
Small oscillations indicate the Monte-Carlo statistical error.
 		\label{gdos4}
	}
\end{figure}

The results of WPIMC simulations of RDFs for fermions with the same spin projection and DOS are presented in Fig.~\ref{gdos3} at $n=0.6 $ and in Fig.~\ref{gdos4} at $n=1$.
The temperature is fixed at $T=60$~K and four different values of $r_s$ are considered. With increasing density the height of the peaks is rising for both hardnesses, but at $n=0.6$ the height is about twice as high as at $n=1$. This difference strongly affects the DOS behavior. In general, for a low density ($r_s=2.3$) the DOS at $n=0.6$ is higher than the one at $n=1$. With increasing density the DOS curve goes down. Then for the highest density $r_s=1.47$
the DOS for $n=0.6$ and $n=1$ practically coincide with each other and are very close to the ideal DOS.

To calculate an RDF, $W(E)$ and DOS Markovian chains of particle configurations were generated using WPIMC (see Supplemental material). The configurations with numbers $10^6 - 3\times 10^6$ for systems of 300, 600 and 900 particles represented by twenty and forty ``beads'' were considered as equilibrium.
We used a standard basic Monte Carlo cell with periodic boundary conditions.
The convergence and statistical error of the calculated functions were tested with increasing number of Monte Carlo steps,
number of particles and beads at a different hardness of the pseudopotential.
It turned out that 600 particles represented by 20 beads were enough to reach the convergence.

With decreasing histogram interval in the energy distribution the statistical error increases due to the worsening of statistics in each interval,
so the compromise between a reasonable width of the discrete interval and the statistical error has to be achieved.
In our simulations the statistical error is of the order of small random oscillations of the distribution functions.

\section{Discussion}
Density of states (DOS) is known to determine the properties of matter and can be used to compute
the thermodynamic properties of a wide variety  systems of particles.
DOSs have been considered and used many times in the literature for calculation of thermodynamic properties of the classical system
\cite{wang2001determining,wang2001efficient,faller2003density,vogel2013generic,liang2005generalized,moreno2022portable,bornn2013adaptive,atchade2010wang}.  
This article deals with the DOS of quantum systems, so the Wigner formulation of quantum mechanics has been used
to derive the new path integral representation of the quantum DOS.
For the $2 {\rm D} $ quantum system of strongly correlated soft--sphere fermions
we present the DOS, internal energy distribution and the spin--resolved RDF obtained by the new
path integral Monte Carlo method (WPIMC) \cite{filinov2022bound} for the hardness of the soft--sphere potential of the order of unity
($n=0.2$, 0.6, 1.0, 1.4). The peculiarities of the dependences of the DOS as a function of
the hardness of the soft--sphere potential and particle density has been investigated and explained.
The WPIMC calculations for greater hadnesses are in progress.

%
%
%
%
%
\section*{Acknowledgements}
We thank G.\,S.~Demyanov for comments and help in numerical aspects.
We value stimulating discussions with Prof. M. Bonitz, T. Schoof, S. Groth and T. Dornheim (Kiel).
The authors acknowledge the JIHT RAS Supercomputer
Centre, the Joint Supercomputer Centre of the Russian
Academy of Sciences, and the Shared Resource Centre Far
Eastern Computing Resource IACP FEB RAS for providing
computing time. 

\subsection*{References.\label{bibby}}
\providecommand{\newblock}{}


\begin{thebibliography}{10}
	\expandafter\ifx\csname url\endcsname\relax
	\def\url#1{{\tt #1}}\fi
	\expandafter\ifx\csname urlprefix\endcsname\relax\def\urlprefix{URL }\fi
	\providecommand{\eprint}[2][]{\url{#2}}
	
	\bibitem{harrison2012electronic}
	Harrison W~A 2012 {\em Electronic structure and the properties of solids: the
		physics of the chemical bond\/} (Courier Corporation)
	
	\bibitem{wang2001determining}
	Wang F and Landau D 2001 {\em Physical Review E\/} {\bf 64} 056101
	
	\bibitem{wang2001efficient}
	Wang F and Landau D~P 2001 {\em Physical review letters\/} {\bf 86} 2050
	
	\bibitem{faller2003density}
	Faller R and de~Pablo J~J 2003 {\em The Journal of chemical physics\/} {\bf
		119} 4405--4408
	
	\bibitem{vogel2013generic}
	Vogel T, Li Y~W, W{\"u}st T and Landau D~P 2013 {\em Physical review letters\/}
	{\bf 110} 210603
	
	\bibitem{liang2005generalized}
	Liang F 2005 {\em Journal of the American Statistical Association\/} {\bf 100}
	1311--1327
	
	\bibitem{moreno2022portable}
	Moreno F, Davis S and Peralta J 2022 {\em Computer Physics Communications\/}
	{\bf 274} 108283
	
	\bibitem{bornn2013adaptive}
	Bornn L, Jacob P~E, Del~Moral P and Doucet A 2013 {\em Journal of Computational
		and Graphical Statistics\/} {\bf 22} 749--773
	
	\bibitem{atchade2010wang}
	Atchad{\'e} Y~F and Liu J~S 2010 {\em Statistica Sinica\/}  209--233
	
	\bibitem{martin2004electronic}
	Martin R 2004 {\em Cambridge Daw MS, Baskes MI (1984) Phys Rev B\/} {\bf
		296443}
	
	\bibitem{seo2014first}
	Seo D~H, Shin H, Kang K, Kim H and Han S~S 2014 {\em The Journal of Physical
		Chemistry Letters\/} {\bf 5} 1819--1824
	
	\bibitem{ma2015machine}
	Ma X, Li Z, Achenie L~E and Xin H 2015 {\em The journal of physical chemistry
		letters\/} {\bf 6} 3528--3533
	
	\bibitem{ratcliff2017challenges}
	Ratcliff L~E, Mohr S, Huhs G, Deutsch T, Masella M and Genovese L 2017 {\em
		Wiley Interdisciplinary Reviews: Computational Molecular Science\/} {\bf 7}
	e1290
	
	\bibitem{galli1996linear}
	Galli G 1996 {\em Current Opinion in Solid State and Materials Science\/} {\bf
		1} 864--874
	
	\bibitem{saad2010numerical}
	Saad Y, Chelikowsky J~R and Shontz S~M 2010 {\em SIAM review\/} {\bf 52} 3--54
	
	\bibitem{goedecker1999linear}
	Goedecker S 1999 {\em Reviews of Modern Physics\/} {\bf 71} 1085
	
	\bibitem{vorontsov2003entropic}
	Vorontsov-Velyaminov P and Lyubartsev A 2003 {\em Journal of Physics A:
		Mathematical and General\/} {\bf 36} 685
	
	\bibitem{lee1993new}
	Lee J 1993 {\em Physical review letters\/} {\bf 71} 211
	
	\bibitem{luyten1971review}
	Luyten W 1971 {\em Journal of the Royal Astronomical Society of Canada\/} {\bf
		65} 304
	
	\bibitem{potekhin2010physics}
	Potekhin A~Y 2010 {\em Physics-Uspekhi\/} {\bf 53} 1235
	
	\bibitem{feynmanquantum}
	Feynman R~P and Hibbs A~R 1965 {\em Quantum Mechanics and Path Integrals\/}
	(New York: McGraw-Hill)
	
	\bibitem{zamalin1977monte}
	Zamalin V, Norman G and Filinov V 1977 The monte carlo method in statistical
	thermodynamics
	
	\bibitem{EbelForFil}
	Ebeling W, Fortov V and Filinov V 2017 {\em Quantum Statistics of Dense Gases
		and Nonideal Plasmas\/} (Berlin: Springer)
	
	\bibitem{ForFilLarEbl}
	Fortov V, Filinov V, Larkin A and Ebeling W 2020 {\em Statistical physics of
		Dense Gases and Nonideal Plasmas\/} (Moscow: PhysMatLit)
	
	\bibitem{dornheim2018uniform}
	Dornheim T, Groth S and Bonitz M 2018 {\em Physics Reports\/} {\bf 744} 1--86
	
	\bibitem{ceperley1995path}
	Ceperley D~M 1995 {\em Reviews of Modern Physics\/} {\bf 67} 279
	
	\bibitem{pollock1984simulation}
	Pollock E~L and Ceperley D~M 1984 {\em Physical Review B\/} {\bf 30} 2555
	
	\bibitem{singer1988path}
	Singer K and Smith W 1988 {\em Molecular Physics\/} {\bf 64} 1215--1231
	
	\bibitem{filinov2022solution}
	Filinov V, Syrovatka R and Levashov P 2022 {\em Molecular Physics\/}  e2102549
	
	\bibitem{ceperley1991fermion}
	Ceperley D~M 1991 {\em Journal of statistical physics\/} {\bf 63} 1237
	
	\bibitem{ceperley1992path}
	Ceperley D~M 1992 {\em Physical review letters\/} {\bf 69} 331
	
	\bibitem{larkin2017pauli}
	Larkin A, Filinov V and Fortov V 2017 {\em Contributions to Plasma Physics\/}
	{\bf 57} 506--511
	
	\bibitem{larkin2017peculiarities}
	Larkin A, Filinov V and Fortov V 2017 {\em Journal of Physics A: Mathematical
		and Theoretical\/} {\bf 51} 035002
	
	\bibitem{wigner1934interaction}
	Wigner E 1934 {\em Physical Review\/} {\bf 46} 1002
	
	\bibitem{Tatr}
	Tatarskii V~I 1983 {\em Soviet Physics Uspekhi\/} {\bf 26} 311
	
	\bibitem{filinov2021monte}
	Filinov V, Levashov P and Larkin A 2021 {\em Journal of Physics A: Mathematical
		and Theoretical\/} {\bf 55} 035001
	
	\bibitem{kubo2012statistical}
	Kubo R, Toda M and Hashitsume N 2012 {\em Statistical physics II:
		nonequilibrium statistical mechanics\/} vol~31 (Springer Science \& Business
	Media)
	
	\bibitem{sese2020real}
	Ses{\'e} L~M 2020 {\em Entropy\/} {\bf 22} 1338
	
	\bibitem{NormanZamalin}
	Zamalin V and Norman G 1973 {\em USSR Computational Mathematics and
		Mathematical Physics\/} {\bf 13} 169--183
	
	\bibitem{LarkinFilinovCPP}
	Larkin A, Filinov V and Fortov V 2016 {\em Contributions to Plasma Physics\/}
	{\bf 56} 187--196
	
	\bibitem{filinov2022bound}
	Filinov V, Larkin A and Levashov P 2022 {\em Universe\/} {\bf 8} 79
	
	\bibitem{filinov2020uniform}
	Filinov V, Larkin A and Levashov P 2020 {\em Physical Review E\/} {\bf 102}
	033203
	
	\bibitem{demyanov2022derivation}
	Demyanov G and Levashov P 2022 {\em arXiv preprint arXiv:2205.09885\/}
	
	\bibitem{Ke63}
	Kelbg G 1963 {\em Ann. Physik\/} {\bf 457} 354
	
	\bibitem{hansen1973statistical}
	Hansen J~P 1973 {\em Physical Review A\/} {\bf 8} 3096
	
	\bibitem{kelbg}
	Ebeling W, Hoffmann H and Kelbg G 1967 {\em Beitr{\"a}ge aus der
		Plasmaphysik\/} {\bf 7} 233--248
	
	\bibitem{afilinov-etal.04pre}
	Filinov A, Golubnychiy V, Bonitz M, Ebeling W and Dufty J 2004 {\em Phys. Rev.
		E\/} {\bf 70} 04641
	
	\bibitem{ebeling_sccs05}
	Ebeling W, Filinov A, Bonitz M, Filinov V and Pohl T 2006 {\em Journal of
		Physics A: Mathematical and General\/} {\bf 39} 4309
	
	\bibitem{KTR94}
	Klakow D, Toepffer C and Reinhard P~G 1994 {\em The Journal of chemical
		physics\/} {\bf 101} 10766--10774
	
	\bibitem{kirkwood1935statistical}
	Kirkwood J~G 1935 {\em The Journal of chemical physics\/} {\bf 3} 300--313
	
	\bibitem{fisher1964statistical}
	Fisher I~Z 1964 {\em Statistical theory of liquids\/} (University of Chicago
	Press)
	
	\bibitem{barker1972theories}
	Barker J and Henderson D 1972 {\em Annual review of physical chemistry\/} {\bf
		23} 439--484
	
	\bibitem{kirkwood1950radial}
	Kirkwood J~G, Maun E~K and Alder B~J 1950 {\em The Journal of Chemical
		Physics\/} {\bf 18} 1040--1047
	
\end{thebibliography}
%
%
%
%

\end{document}